\documentclass[12pt]{article}
\usepackage{amsmath,amsfonts,latexsym,amsthm,amssymb}
\newcommand{\K}{\mathcal K}

\DeclareMathOperator{\I}{Im}
\DeclareMathOperator{\R}{Re}
\begin{document}
\newtheorem{teo}{Theorem}
\title{Distributed Order Derivatives and Relaxation Patterns}
\author{Anatoly N. Kochubei
\\ \footnotesize Institute of Mathematics,\\
\footnotesize National Academy of Sciences of Ukraine,\\
\footnotesize Tereshchenkivska 3, Kiev, 01601 Ukraine}
\date{}
\maketitle
\vspace*{2cm}
Running head:\quad  ``Distributed Order Derivatives''

\vspace*{2cm}
\begin{abstract}
We consider equations of the form
$$
\left( \mathbb D_{(\rho )}u\right) (t)=-\lambda u(t),
\quad t>0,
$$
where $\lambda >0$, $\mathbb D_{(\rho )}$ is a distributed order derivative, that is
$$
\mathbb D_{(\rho )}\varphi (t)=\int\limits_0^1 (\mathbb D^{(\alpha )}\varphi )(t)\,d\rho (\alpha ),
$$
$\mathbb D^{(\alpha )}$ is the Caputo-Dzhrbashyan fractional
derivative of order $\alpha$, $\rho$ is a positive measure.
\par
The above equation is used for modeling anomalous, non-exponential
relaxation processes. In this work we study asymptotic behavior of solutions of the above equation, depending on properties of the measure $\rho$.

\end{abstract}

\vspace{2cm}

\newpage
\section{INTRODUCTION}

Anomalous, non-exponential, relaxation processes occur in various branches of physics; see e.g. \cite{BGJ,MK,T} and references therein. Just as the exponential function $e^{-\lambda t}$ ($\lambda >0$) appearing in the description of classical relaxation is a solution of the simplest differential equation $u'=-\lambda u$ (with the initial condition $u(0)=1$), new kinds of derivatives are used to obtain models of slow relaxation.

It is generally accepted now that the power law of relaxation corresponds to the Cauchy problem
\begin{equation}
\left( \mathbb D^{(\alpha )}u\right) (t)=-\lambda u(t),\ t>0;\quad u(0)=1,
\end{equation}
where
\begin{equation}
\left( \mathbb D^{(\alpha )}u\right) (t)=\frac{1}{\Gamma (1-\alpha )}
\left[ \frac{d}{dt}\int\limits_0^t(t-\tau )^{-\alpha }u(\tau )\,d\tau -t^{-\alpha }u(0)\right]
\end{equation}
is the Caputo-Dzhrbashyan fractional derivative of order $\alpha \in (0,1)$; we refer to \cite{EIK,KST} for various notions and results regarding fractional differential equations. The solution of the problem (1) has the form $u(t)=E_\alpha (-\lambda t^\alpha )$ where $E_\alpha$ is the Mittag-Leffler function, and $u(t)\sim Ct^{-\alpha }$ (here and below we denote various positive constants by the same letter $C$), as $t\to \infty$. This kind of evolution describes the temporal behavior related to $\alpha$-fractional diffusion typical for fractal media.

A still slower, logarithmic relaxation \cite{GM2,MMGS} is described by the Cauchy problem
\begin{equation}
\left( \mathbb D^{(\mu )}u\right) (t)=-\lambda u(t),\ t>0;\quad u(0)=1,
\end{equation}
where
\begin{equation}
\left( \mathbb D^{(\mu )}u\right) (t)=\int\limits_0^1\left( \mathbb D^{(\alpha )}u\right) (t)\mu (\alpha )\,d\alpha ,
\end{equation}
$\mu$ is a non-negative continuous function on $[0,1]$. A rigorous mathematical treatment of the problem (3) for general classes of the weights $\mu$ was given in \cite{K1,K2}. Some nonlinear equations with distributed order derivatives were studied in \cite{AOP}. As above, this kind of relaxation models is connected with models of ultraslow diffusion \cite{CGSG,GM1,H,K1,MS,UG}; these papers contain further references on related subjects.

Let $u_\lambda (t)$ be the solution of the problem (3) (the notation is changed slightly, compared to \cite{K1}). If $\mu (0)\ne 0$, then
\begin{equation}
u_\lambda (t)\sim C(\log t)^{-1},\quad t\to \infty .
\end{equation}
If $\mu (\alpha )\sim a\alpha^\nu$, $\alpha \to 0$ ($a>0$, $\nu >0$), then
\begin{equation}
u_\lambda (t)\sim C(\log t)^{-1-\nu },\quad t\to \infty .
\end{equation}

It was assumed everywhere in \cite{K1} that $\mu \in C^3[0,1]$ and $\mu (1)\ne 0$; in fact, in the investigation of the asymptotic behavior of $u_\lambda$ we use only that $\mu \in L_1(0,1)$. Therefore the arguments in \cite{K1} cover the case where
$$
\mu (\alpha )\sim a\alpha^{-\nu}, \quad \alpha \to 0,
$$
with $a>0$, $0<\nu <1$, and yield the asymptotics
\begin{equation}
u_\lambda (t)\sim C(\log t)^{-1+\nu },\quad t\to \infty .
\end{equation}

In all the above cases, the exponential function $e^{-\lambda t}$, the Mittag-Leffler powerlike evolution $E_\alpha (-\lambda t^\alpha )$, and all the logarithmic evolutions (5)-(7), the resulting functions are completely monotone, that is $(-1)^j u_\lambda^{(j)}(t)\ge 0$, $j=0,1,2,\ldots$, for all $t$.

In this paper we look for other possible relaxation patterns corresponding to the weight functions $\mu$ tending to 0 (at the origin) faster than the power function (Section 2) or to the definition of the distributed order derivative not by the formula (4) but by the expression
\begin{equation}
\left( \mathbb D_{(\rho )}u\right) (t)=\int\limits_0^1 \left( \mathbb D^{(\alpha )}u\right) (t)\,d\rho (\alpha ),
\end{equation}
where $\rho$ is a jump measure (Section 3). In these cases the solutions remain completely monotone while their asymptotic behavior can be quite diverse, from the iterated logarithmic decay to a function decaying faster than any power of logarithm but slower than any power function.

\section{THE MAIN CONSTRUCTIONS}

Suppose that $\rho$ is a positive finite measure on $[0,1]$ not concentrated at 0, and consider the distributed order derivative (8). Substituting (2) into (8) we find that
\begin{equation}
\left( \mathbb D_{(\rho )}u\right) (t)=\frac{d}{dt}\int\limits_0^t k(t-\tau )u(\tau )\,d\tau -k(t)u(0)
\end{equation}
where
\begin{equation}
k(s)=\int\limits_0^1 \frac{s^{-\alpha }}{\Gamma (1-\alpha )}d\rho (\alpha ),\quad s>0.
\end{equation}
The right-hand side of (9) makes sense for a continuous function $u$, for which the derivative $\dfrac{d}{dt}\int\limits_0^t k(t-\tau )u(\tau )\,d\tau$ exists.

It is clear from (10) that $k\in L_1^{\text{loc}}(0,\infty )$, and $k$ is decreasing. Therefore the function $k$ possesses the Laplace transform
$$
\K (p)=\int\limits_0^\infty k(s)e^{-ps}\,ds=\int\limits_0^1 p^{\alpha -1}d\rho (\alpha ),\quad \R p>0.
$$
The holomorphic function $\K (p)$ can be extended analytically onto the whole complex plane cut along the half-axis $\mathbb R_-=\{ \I p=0,\R p\le 0\}$. Obviously, $\K (p)\to 0$, as $|p|\to \infty$ (a precise asymptotics is found for some cases in \cite{K1} and Section 3 below).

Note that if we consider the moments of the measure $\rho$,
$$
M_n=\int\limits_0^1 x^n\,d\rho (x),
$$
and introduce their generating function
$$
M(z)=\sum\limits_{n=0}^\infty M_n\frac{z^n}{n!}=\int\limits_0^1 e^{xz}\,d\rho (x),
$$
then $\K (p)=p^{-1}M(\log p)$. The moment generating functions were studied by a number of authors (for example, \cite{Curt,RRS}).

Considering the relaxation equation
\begin{equation}
\left( \mathbb D_{(\rho )}u\right) (t)=-\lambda u(t),
\quad t>0,
\end{equation}
with $\lambda >0$, we apply formally the Laplace transform (which is justified post factum, using the smoothness properties and the asymptotic behavior of the solution).

For the Laplace transform $\widetilde{u_\lambda }(p)$ of the solution $u_\lambda (t)$ of the equation (11) satisfying the initial condition $u_\lambda (0)=1$ we get the expression
\begin{equation}
\widetilde{u_\lambda }(p)=\frac{\K (p)}{p\K (p)+\lambda }.
\end{equation}
Since $p\K (p)\to \infty$, as $p\to \infty$, we have $\widetilde{u_\lambda }(p)\sim p^{-1}$, $p=\sigma +i\tau$, $\sigma ,\tau \in \mathbb R$, $|\tau |\to \infty$. Therefore \cite{DP}, $\widetilde{u_\lambda }$ is the Laplace transform of some function $u_\lambda (t)$, and for almost all t,
\begin{equation}
u_\lambda (t)=\frac{d}{dt}\frac{1}{2\pi i}\int\limits_{\gamma
-i\infty}^{\gamma +i\infty }\frac{e^{pt}}{p}\frac{\K (p)}{p\K (p)+\lambda }\,dp.
\end{equation}

Note that for $p\in \mathbb C\setminus \mathbb R_-$ we have
$$
\I p\K (p)=\int\limits_0^1|p|^\alpha \sin (\alpha
\arg p)\,d\rho (\alpha ),
$$
so that $\I p\K (p)=0$ only for $\arg p=0$. This means that $p\K (p)+\lambda \ne 0$, and the representation (13) is valid for an arbitrary $\gamma >0$.

As in \cite{K1}, in the present more general situation we deform the contour of integration, and then differentiate under the integral, so that
\begin{equation}
u_\lambda (t)=\frac{1}{2\pi i}\int\limits_{S_{\gamma ,\omega}}
e^{pt}\frac{\K (p)}{p\K (p)+\lambda }\,dp
\end{equation}
where the contour $S_{\gamma ,\omega}$ consists of the arc
$$
T_{\gamma ,\omega }=\{ p\in \mathbb C:\ |p|=\gamma, |\arg p|\le
\omega \pi \},\quad \frac12 <\omega <1,
$$
and two rays
$$
\Gamma_{\gamma ,\omega }^\pm =\{ p\in \mathbb C:\ |\arg p|=\pm \omega \pi , |p|\ge \gamma \}.
$$

Just as in the case of a measure $\rho$ with a smooth density considered in \cite{K1}, it is easy to show that the function $u_\lambda$ belongs to $C^\infty (0,\infty )$ and is continuous at the origin; from its construction and the formula (9), it follows that the initial condition $u_\lambda (0)=1$ is indeed satisfied.

Let us consider first the case of a measure $\rho$ with a continuous density, $d\rho (\alpha )=\mu (\alpha )\,d\alpha $. This case was investigated in \cite{K1}, and it was proved that $u_\lambda$ {\it is completely monotone} (various additional assumptions made in \cite{K1} and needed for other problems studied in that paper, were not actually used here).

In this paper we consider the case of a different behavior of the density $\mu$ near the origin, implying a different asymptotics of $u_\lambda (t)$, as $t\to \infty$.

\medskip
\begin{teo}
If $\mu \in C[0,1]$, and
$$
\mu (\alpha )\sim a\alpha^\gamma e^{-\frac{\beta }{\alpha }},\quad \text{as $\alpha \to 0$},
$$
where $a>0$, $\gamma >-1$, $\beta >0$, then
\begin{equation}
u_\lambda (t)\sim C(\log t)^{-\frac{\gamma }2-\frac34}e^{-2\sqrt{\beta }(\log t)^{\frac12}},\quad t\to \infty .
\end{equation}
\end{teo}

\medskip
{\it Proof}. Let us write $\K (p)$ as
$$
\K (p)=p^{-1}\int\limits_0^\infty e^{-\alpha z}\mu_1(\alpha )\,d\alpha ,\quad z=\log \frac1p,
$$
where $\mu_1$ is the extension of $\mu$ by zero onto $\mathbb R_+$, and use an asymptotic result for Laplace integrals from \cite{Ri} (Theorem 13.1, Case 9). We get
$$
\K (p)\sim 2ap^{-1}\left( \frac{\beta}z\right)^{\frac{\gamma +1}2}K_{\gamma +1}(2\sqrt{\beta z}),\quad p\to +0,
$$
where, as before, $z=\log \frac1p$ ($\to \infty$), $K_m$ is the McDonald function. It is well known that $K_m(t)\sim \left( \frac{\pi }2\right)^{1/2}t^{-1/2}e^{-t}$, $t\to \infty$, so that
\begin{equation}
\K (p)\sim Cp^{-1}L(\frac1p ), \quad p\to +0,
\end{equation}
where
$$
L(s)=(\log s)^{-\frac{\gamma }2-\frac34}e^{-2\sqrt{\beta }(\log s)^{\frac12}}.
$$

It follows from (16) (or directly from the definition of $\K (p)$) that $p\K (p)\to 0$, as $p\to +0$. Therefore by (12),
$$
\widetilde{u_\lambda }(p)\sim \lambda^{-1}\K (p),\quad \text{as $p\to +0$}.
$$
Since we already know that the function $u_\lambda$ is monotone, we may apply the Karamata-Feller Tauberian theorem (see Chapter XIII in \cite{Fe}) which implies the desired asymptotics of $u_\lambda (t)$, $t\to \infty$. $\qquad \blacksquare$

\medskip
In the case under consideration, the function $u_\lambda (t)$ decreases at infinity slower than any negative power of $t$, but faster than any negative power of $\log t$. It is also seen from (15) that the decrease is accelerated if $\beta >0$ becomes bigger, so that the less the weight function $\mu$ is near 0, the faster is the relaxation for large times.

\section{THE STEP STIELTJES WEIGHT}

Let us consider the case where the integral in (8) is a Stieltjes integral corresponding to a right continuous non-decreasing step function $\rho (\alpha )$. In order to investigate a sufficiently general situation, we assume that the function $\rho$ has two sequences of jump points, $\beta_n$ and $\nu_n$, $n=0,1,2,\ldots$, where $\beta_n\to 0$, $\nu_n\to 1$, $\beta_0=\nu_0\in (0,1)$. We may assume that the sequence $\{ \beta_n\}$ is strictly decreasing while $\{ \nu_n\}$ is strictly increasing.

Denote $\Delta \rho (t)=\rho (t)-\rho (t-0)$, $\xi_n=\Delta \rho (\beta_n)$ ($n\ge 0$), $\eta_n=\Delta \rho (\nu_n)$ ($n\ge 1$); we have $\xi_n,\eta_n>0$ for all $n$. It will be convenient to assume that $\beta_0<e^{-1}$ and to
write $\eta_0=0$. Since $\rho$ is a finite measure, we have also
\begin{equation}
\sum\limits_{n=0}^\infty \xi_n <\infty ,\quad \sum\limits_{n=0}^\infty \eta_n <\infty .
\end{equation}

By (10),
$$
k(s)=\sum\limits_{n=0}^\infty \frac{\xi_n}{\Gamma
(1-\beta_n)}s^{-\beta_n}+\sum\limits_{n=1}^\infty \frac{\eta_n}{\Gamma
(1-\nu_n)}s^{-\nu_n}, \quad s>0,
$$
so that
\begin{equation}
\K (p)=\sum\limits_{n=0}^\infty \xi_np^{\beta_n-1}+\sum\limits_{n=1}^\infty \eta_np^{\nu_n-1}.
\end{equation}

As before, we denote by $u_\lambda (t)$ the solution of the relaxation equation (11) with the initial condition $u_\lambda (0)=1$. The symbol $f\asymp g$ will, as usual, mean that $f=O(g)$ and $g=O(f)$.

\medskip
\begin{teo}
$\mathrm{(i)}$ The function $u_\lambda$ is completely monotone.

\begin{equation}
\mathrm{(ii)}\quad u_\lambda (x)\asymp  \sum\limits_{n=0}^\infty \left[ \frac{\xi_n}{\Gamma (2-\beta_n)}x^{-\beta_n}+\frac{\eta_n}{\Gamma
(2-\nu_n)}x^{-\nu_n}\right] ,\quad x\to \infty ;
\end{equation}

$\mathrm{(iii)}$ If $\sum\limits_{n=0}^\infty \xi_n \left( \log \log \frac1{\beta_n}\right)^b<\infty$ ($b>0$), then
\begin{equation}
u_\lambda (x)=O\left( \frac1{(\log \log x)^b}\right),\quad x\to \infty. \end{equation}

$\mathrm{(iv)}$ If $\sum\limits_{n=0}^\infty \xi_n\beta_n^{-b}<\infty$ ($b>0$), then
\begin{equation}
u_\lambda (x)=O\left( \frac1{(\log x)^b}\right),\quad x\to \infty. \end{equation}
\end{teo}

\medskip
{\it Proof}. Using the representation (14) and following \cite{K1} we can write
\begin{multline*}
u_\lambda (t)=\frac{1}{2\pi i}\int\limits_{T_{\gamma ,\omega}}
e^{pt}\frac{\K (p)}{p\K (p)+\lambda }\,dp+
\frac{1}\pi \I \int\limits_\gamma^\infty r^{-1}e^{tre^{i\omega \pi
}}\,dr\\
-\frac{\lambda }\pi \I \int\limits_\gamma^\infty \frac{e^{tre^{i\omega \pi
}}}{r\left( re^{i\omega \pi }\K (re^{i\omega \pi
})+\lambda \right)}\,dr\overset{\text{def}}{=}J_1+J_2-J_3.
\end{multline*}
Let us substantiate passing to the limit as $\gamma\to 0$.

As $p\to 0$, $p\K (p)\to 0$, so that
$$
\left| \frac{\K (p)}{p\K (p)+\lambda }\right| \le C\sum\limits_{n=0}^\infty \left( \xi_np^{\beta_n-1}+\eta_np^{\nu_n-1}\right) ,
$$
whence
$$
|J_1|\le Ce^{\gamma t}\sum\limits_{n=0}^\infty \left( \xi_n\gamma^{\beta_n}+\eta_n\gamma^{\nu_n}\right) \to 0,
$$
as $\gamma\to 0$. It was shown in \cite{K1} that
$$
J_2\longrightarrow -\frac{1}\pi \int\limits_0^\infty s^{-1}e^{-s}
\sin (s\tan \omega \pi )\,ds,
$$
as $\gamma\to 0$.

The integral $J_3$ is the sum of
$$
I_1=\frac{\lambda }\pi \int\limits_\gamma^\infty \I \left( \frac{e^{tre^{i\omega \pi
}}}{r}\right) \R \left( \frac{1}{re^{i\omega \pi }\K (re^{i\omega \pi
})+\lambda }\right)\,dr
$$
and
$$
I_2=\frac{\lambda }\pi \int\limits_\gamma^\infty \R \left( \frac{e^{tre^{i\omega \pi
}}}{r}\right) \I \left( \frac{1}{re^{i\omega \pi }\K (re^{i\omega \pi
})+\lambda }\right)\,dr.
$$We have
$$
\I \left( \frac{e^{tre^{i\omega \pi }}}{r}\right) =r^{-1}e^{tr\cos
\omega \pi}\sin (tr\sin \omega \pi ),
$$
and this expression has a finite limit, as $r\to 0$. Since also
$p\K (p)\to 0$, as $p\to 0$, we see that we may pass to the limit in
$I_1$, as $\gamma \to 0$.

Let
$$
\Phi (r,\omega )=\I \frac{1}{re^{i\omega \pi }\K (re^{i\omega \pi
})+\lambda }.
$$
Substituting (18) and denoting
\begin{multline*}
G(r,\omega )=\left\{ \sum\limits_{n=0}^\infty \left[ \xi_nr^{\beta_n}\cos (\omega \pi \beta_n)+\eta_nr^{\nu_n}\cos (\omega \pi \nu_n)\right]+\lambda \right\}^2\\
+\left\{ \sum\limits_{n=0}^\infty \left[ \xi_nr^{\beta_n}\sin (\omega \pi \beta_n)+\eta_nr^{\nu_n}\sin (\omega \pi \nu_n)\right]\right\}^2
\end{multline*}
we find that
$$
\Phi (r,\omega )=-\frac{\sum\limits_{n=0}^\infty \left[ \xi_nr^{\beta_n}\sin (\omega \pi \beta_n)+\eta_nr^{\nu_n}\sin (\omega \pi \nu_n)\right]}{G(r,\omega )}.
$$
The denominator tends to $\lambda^2$, as $r\to 0$. Noting that
$$
\R \left( \frac{e^{tre^{i\omega \pi }}}{r}\right) =r^{-1}e^{tr\cos \omega \pi }\cos (tr\sin \omega \pi )
$$
and using (17) we find that the integrand in $I_2$ belongs to $L_1(0,\infty )$.

Passing to the limit $\gamma \to 0$ we obtain the representation
\begin{multline*}
u_\lambda (t)=-\frac{1}\pi \int\limits_0^\infty s^{-1}e^{-s}\sin
(s\tan \omega \pi )\,ds -\frac{\lambda }\pi \int\limits_0^\infty
r^{-1}e^{tr\cos \omega \pi }\sin (tr\sin \omega \pi )\Psi
(r,\omega )\,dr \\
-\frac{\lambda }\pi \int\limits_0^\infty
r^{-1}e^{tr\cos \omega \pi }\cos (tr\sin \omega \pi )\Phi
(r,\omega )\,dr
\end{multline*}
where
$$
\Psi (r,\omega )=\frac{\sum\limits_{n=0}^\infty \left[ \xi_nr^{\beta_n}\cos (\omega \pi \beta_n)+\eta_nr^{\nu_n}\cos (\omega \pi \nu_n)\right]+\lambda }{G(r,\omega )}.
$$

Next, let us pass to the limit, as $\omega \to 1$. It follows from the Lebesgue theorem that the first two integrals tend to zero, and we get
$$
u_\lambda (t)=\frac{\lambda }\pi \int\limits_0^\infty
r^{-1}e^{-tr}\frac{\sum\limits_{n=0}^\infty \left[ \xi_nr^{\beta_n}\sin (\pi \beta_n)+\eta_nr^{\nu_n}\sin (\pi \nu_n)\right]}{G(r,1)}\,dr,
$$
that is, up to a positive factor, $u_\lambda$ is the Laplace transform of a locally integrable non-negative function bounded at infinity. Therefore $u_\lambda$ is completely monotone.

\medskip
$\mathrm{(ii)}$ It will be convenient to turn to the Laplace-Stieltjes transform instead of the Laplace transform. Set
$$
\varkappa (x)=\int\limits_0^xk(s)\,ds,\quad v_\lambda (x)=\int\limits_0^xu_\lambda (s)\,ds,
$$
so that the Laplace-Stieltjes transforms are as follows:
$$
\widehat{\varkappa }(p)=\int\limits_0^\infty e^{-px}\,d\varkappa (x)=\K (p),\quad \widehat{v_\lambda }(p)=\widetilde{u_\lambda }(p).
$$
By (12), $\widehat{v_\lambda }(p)=l(p)\widehat{\varkappa }(p)$ where $l(p)=\dfrac1{p\K (p)+\lambda }$ is a slowly varying function near the origin.

We have
\begin{equation}
\varkappa (x)=\sum\limits_{n=0}^\infty \frac{\xi_n }{\Gamma (2-\beta_n)}x^{1-\beta_n}+\sum\limits_{n=1}^\infty \frac{\eta_n }{\Gamma (2-\nu_n)}x^{1-\nu_n},\quad x>0.
\end{equation}
The function $\varkappa$ is monotone increasing. Denote
$$
\varkappa^*(\zeta )=\limsup\limits_{x\to \infty }\frac{\varkappa (\zeta x)}{\varkappa (x)},\quad \zeta >1.
$$

Since $\beta_n,\nu_n\in (0,1)$, we find that
\begin{equation}
\varkappa (\zeta x)\le \zeta \varkappa (x),
\end{equation}
so that $\varkappa^*(\zeta )\le \zeta$. Thus, $\varkappa$ is an O-regularly varying function (see \cite{BGT}, especially Corollary 2.0.6).

On the other hand, it is seen from (22) that $\varkappa^*(\zeta )\ge 1$ for $\zeta >1$, so that $\varkappa^*(+1)=1$. Thus we are within the conditions of the ``Ratio Tauberian Theorem'' (\cite{BGT}, Theorem 2.10.1), which yields the relation $v_\lambda (x)\sim l(\frac1x)\varkappa (x)$, $x\to \infty$, so that
\begin{equation}
v_\lambda (x)\sim C\varkappa (x), \quad x\to \infty .
\end{equation}

In order to pass from (24) to (19), we have to check further Tauberian conditions. Since $u_\lambda$ is completely monotone, it is, in particular, non-increasing, thus belonging to the class BI (see Section 2.2 in \cite{BGT} for the definitions of this class and the class PI used below). It follows from (23) that also $\varkappa \in \text{BI}$.

Next,
$$
\varkappa (x)\ge \sum\limits_{n=0}^\infty \frac{\xi_n }{\Gamma (2-\beta_n)}x^{1-\beta_n}+\frac{\eta_1}{\Gamma (2-\nu_1)}x^{1-\nu_1}.
$$
Since $\zeta^{1-\beta_n}\ge \zeta^{1-\nu_1}$ for $\zeta >1$, we have
\begin{multline*}
\frac{\varkappa (\zeta x)}{\varkappa (x)}\ge \zeta^{1-\nu_1}
\frac{\sum\limits_{n=0}^\infty \frac{\xi_n }{\Gamma (2-\beta_n)}x^{1-\beta_n}+\frac{\eta_1}{\Gamma (2-\nu_1)}x^{1-\nu_1}}
{\sum\limits_{n=0}^\infty \left( \frac{\xi_n }{\Gamma (2-\beta_n)}x^{1-\beta_n}+\frac{\eta_n }{\Gamma (2-\nu_n)}x^{1-\nu_n}\right)} \\
=\zeta^{1-\nu_1}\left\{ 1-\frac{\sum\limits_{n=2}^\infty \frac{\eta_n}{\Gamma (2-\nu_n)}x^{1-\nu_n}}
{\sum\limits_{n=0}^\infty \left( \frac{\xi_n }{\Gamma (2-\beta_n)}x^{1-\beta_n}+\frac{\eta_n }{\Gamma (2-\nu_n)}x^{1-\nu_n}
\right)} \right\}
\end{multline*}
where
$$
\frac{\sum\limits_{n=2}^\infty \frac{\eta_n}{\Gamma (2-\nu_n)}x^{1-\nu_n}}
{\sum\limits_{n=0}^\infty \left( \frac{\xi_n }{\Gamma (2-\beta_n)}x^{1-\beta_n}+\frac{\eta_n }{\Gamma (2-\nu_n)}x^{1-\nu_n}
\right)}
\le Cx^{-1+\nu_1}\sum\limits_{n=2}^\infty \frac{\eta_n}{\Gamma (2-\nu_n)}x^{1-\nu_n}\longrightarrow 0,
$$
as $x\to \infty$. Thus, $\varkappa \in \text{PI}$.

Now we are within the conditions of the O-version of Monotone Density Theorem (\cite{BGT}, Proposition 2.10.3), which implies the required asymptotic relation (19).

$\mathrm{(iii)-(iv)}$ Let us prove (20). The proof of (21) is similar and simpler, and we leave it to a reader. It is obviously sufficient to deal with the first summand in each element of the series in (19).

Let us consider the function
\begin{equation}
\varphi (x)=x^{-a}(\log \log x)^b, \quad x\ge e,
\end{equation}
where $a,b>0$. We have
$$
\varphi' (x)=x^{-a}(\log \log x)^{b-1}\psi (x)
$$
where $\psi (x)=\frac{b}{\log x}-a\log \log x$. It is easy to check that $\psi$ decreases on $[e,\infty )$, $\psi (e)=b$, $\psi (x)\to -\infty$, as $x\to \infty$. The maximal value of the function $\varphi$ is attained at a single point $x_0$ where $\psi (x_0)=0$, that is
$$
\frac{b}{\log x_0}=a\log \log x_0.
$$

Denote $y_0=\log x_0$. Then $\dfrac{b}{y_0}=a\log y_0$, so that $y_0\log y_0=\dfrac{b}a$. We will in fact need an asymptotic behavior of $y_0$ as a function of $a$, as $a\to 0$. It is known (see Section I.5.2 in \cite{Fed}) that
$$
y_0=\log \frac{b}a-\log \log \frac{b}a+O\left( \frac{\log \log \frac{b}a}{\log \frac{b}a}\right) .
$$
Therefore
\begin{equation}
C_1\frac{b}a\left( \log \frac{b}a\right)^{-1}\le x_0\le C_2\frac{b}a\left( \log \frac{b}a\right)^{-1}
\end{equation}
where the constants do not depend on $a,b$.

We have $\varphi (x)\le \varphi (x_0)$, so that, by (25),
$$
x^{-a}\le \frac{\varphi (x_0)}{(\log \log x)^b}.
$$
We need an estimate of $\varphi (x_0)$ making explicit its dependence on $a$. By (26),
$$
x_0^{-a}\le C_3\left( \frac{b}a\right)^{-a}\left( \log \frac{b}a\right)^a
$$
where $\left( \dfrac{b}a\right)^{-a}=e^{a\log \frac{b}a}\to 1$ and $\left( \log \frac{b}a\right)^a=e^{a\log \log \frac{b}a}\to 1$, as $a\to 0$. Next, $x_0\le \dfrac{C_4}a$, so that
$$
\log\log x_0\le \log (\log C_4+\log \frac1{a})\le \log (C_5\log \frac1{a})\le C_6\log \log \frac1{a}
$$
whence $\varphi (x_0)\le C_7(\log \log \frac1{a})^b$.

As a result,
$$
\sum\limits_{n=0}^\infty \frac{\xi_n }{\Gamma (2-\beta_n)}x^{-\beta_n}
\le C_8\left[ \sum\limits_{n=0}^\infty \xi_n \left( \log\log \frac1{\beta_n}\right)^b\right] \frac1{(\log \log x)^b},
$$
and we have proved (20). $\qquad \blacksquare$

\end{document}